"Favoring my playmate seems fair": Inhibitory control and theory of mind in preschoolers' self-disadvantaging behaviors

Dongjie Xie

    School of Psychological and Cognitive Sciences, and

    Beijing Key Laboratory of Behavior and Mental Health,

    Peking University, Beijing, 100871

Meng Pei

    School of Psychological and Cognitive Sciences, and

    Beijing Key Laboratory of Behavior and Mental Health,

    Peking University, Beijing, 100871

Yanjie Su (corresponding author)

    School of Psychological and Cognitive Sciences, and

    Beijing Key Laboratory of Behavior and Mental Health,

    Peking University, Beijing, 100871

    Email: yjsu@pku.edu.cn

**Highlights**

- With interest involved, self-disadvantageous inequity in preschoolers rose with age.
- The improvement of inhibitory control explained this developmental trajectory.
- Better theory of mind was associated with self-disadvantaging tendency.

**Abstract**

The purpose of this study was to investigate the relationship between preschoolers' cognitive abilities and their fairness-related allocation behaviors in a dilemma of equity-efficiency conflict. Four- to 6-year-olds in Experiment 1 ($N = 99$) decided how to allocate 5 reward bells. In the first-party condition, preschoolers were asked to choose among giving more to self (self-advantageous inequity), wasting one bell (equity) or giving more to other (self-disadvantageous inequity); while in the third-party condition, they chose to allocate the extra bell to one of two equally deserving recipients or to waste it. Results showed that compared to the pattern of decision in the third-party condition, preschoolers in the first-party condition were more likely to give the extra bell to other (self-disadvantaging behaviors), and age, inhibitory control (IC) and theory of mind (ToM) were positively correlated with their self-disadvantaging choices, but only IC mediated the relationship between age and self-disadvantaging behaviors. Experiment 2 ($N = 41$) showed that IC still predicted preschoolers' self-disadvantaging behaviors when they could choose only between equity and disadvantageous inequity. These results suggested that IC played a critical role in the implementation of self-disadvantaging behaviors when this required the control over selfishness and envy.

**Acknowledgements**

This work was supported by the National Natural Science Foundation of China (Project Nos. 31571134, 31872782). We thank Ms. Siyuan Shang, Ms. Xiaonan Wang and Mr. Dawei Chen for helpful comments on earlier versions of the manuscript; the teachers and staff in Harvest bilingual kindergartens, Ms. Qichen Wang, Ms. Cui Zhu and Ms. Yiling Wu for their help with data collection.

# Introduction

Humans have regarded fairness as central to morality and social norms (Decety & Wheatley, 2015). Resource allocation is a crucial context to study humans' understanding of fairness and its ontogeny (Tomasello & Vaish, 2013), and equality (everyone receives the same) is a simple rule of fairness (Baumard, Mascaro, & Chevallier, 2012; Rawls, 1971). As children grew older, they would take more factors such as effort into consideration and behave more adult-like (Piaget, 1932; see Hook & Cook, 1979 for a review). Thus, humans would distribute resources according to more complex rules like the equity principle (one receives what is proportional to contribution) (Deutsch, 1975). Fehr and Schmidt (1999) proposed the inequity aversion model as an explanation for humans' fair behaviors, which challenged the standard self-interest model in economics. In addition to equality and equity, efficiency is another important principle in resource allocation. An efficient distribution is considered just (Rawls, 1971).

Recent research revised the inequity aversion model by suggesting that individuals were not upset about inequity per se but were instead upset about the way it was created (Shaw, Choshen-Hillel, & Caruso, 2016). In real-life situations, however, resources cannot always be allocated equitably and individuals may encounter situations where equity is in conflict with efficiency (Choshen-Hillel, Shaw, & Caruso, 2015; Hsu, Anen, & Quartz, 2008; Okun, 1975). An example features a lab manager allocating several old computers and a powerful new one to graduate students. Any efficient plan would be making use of the new one, but doing so would sacrifice equity as the new one gives its user advantage over others (Choshen-Hillel et al., 2015).

To make a relatively satisfactory allocation in such an allocation dilemma, one should be able to anticipate others' social preferences before making a decision. For instance, inferring others' emotional aversion for selfishness encourages us to conform to social norms. Consider a scene where resources could not be allocated equitably. First-party agents had to choose among giving more to self, wasting some to achieve equity and giving more to other, and previous research revealed that adults tended to give more to other which resulted in disadvantageous inequity but promoted efficient resource allocation (Choshen-Hillel et al., 2015). Similarly, third-party agents had to choose to give more to one of the two equally deserving recipients or to waste some; instead of prioritizing efficiency, adults were more likely to make an equitable allocation to avoid dissatisfaction from the recipients for the allocation decision (Choshen-Hillel et al., 2015). A developmental study showed that this self-disadvantaging tendency increased as children grew older (Shaw et al., 2016). It has yet to be explained what specific cognitive abilities enable children to perform self-disadvantaging behaviors. Thus, the current study was conducted mainly to examine the cognitive mechanisms underlying this developmental trajectory in the preschool period. By studying the potential cognitive components, it could provide insights into how humans develop to become social beings, especially in a period with rapid development in social behaviors such as the preschool period (Feldman, 2015). Findings on the cognitive mechanisms would also inform intervention to facilitate children's moral development like training programs for the fairness-related cognitive components.

*From equity to self- disadvantageous inequity*

As children grew older, their knowledge of fairness accumulated. For example, Cooley and Killen (2015) found that 3.5- to 6-year-old children would make negative judgment for the allocator who gave an unequal allocation, even when it benefited the group; and older children took a fairness perspective more when they gave reasons for their negative judgment, such as "it's just not fair to give them less". Further, older children's actual behaviors were more consistent with their knowledge of equity as they were more likely than younger children to propose equitable allocations (see Hook & Cook, 1979 for a review; also see Blake, McAuliffe, & Warneken, 2014) and reject inequity (Blake & McAuliffe, 2011).

However, inequity could be perceived fair if it has legitimate reasons such as "A contributed more than B" (merit-based), "A needs more than B" (need-based), etc. (Feinberg, 1974; Rawls,1971). Adults view inequality or inequity as acceptable if it is achieved by an impartial procedure (Tyler, 2000). Children as young as 5 years old could understand procedural justice and accept an inequitable distribution made through an impartial lottery wheel (Grocke, Rossano, & Tomasello, 2015). But when allocating resources, an impartial procedure such as coin toss is not always appropriate or available (Keren & Teigen, 2010). When rewards could not be allocated equitably between two equally deserving recipients, adults (Choshen-Hillel et al., 2015; Gordon-Hecker, Rosensaft-Eshel, Pittarello, Shalvi, & Bereby-Meyer,

2017) and older children (Shaw & Olson, 2012) as third-parties were more likely than younger ones to waste one reward for equity; but this tendency to waste resource became significantly weaker when the decision would influence their own benefit (Choshen-Hillel et al., 2015; Shaw et al., 2016). If children's self-interest was involved, older ones were more inclined than younger ones to make self-disadvantageous allocations to avoid wasting, i.e., a 1:2 allocation between self and other as opposed to a 1:1 one with the extra reward thrown away (Shaw et al., 2016). That is, inequity that would disadvantage participants was perceived fair if it was created by participants themselves instead of by third parties in the dilemma where equity was in conflict with efficiency (Choshen-Hillel et al., 2015). These findings indicate that children not only accept inequitable yet fair allocations, but even also actively make inequitable allocations if inequity does not entail unfairness (Shaw et al., 2016). In other words, people are not averse to inequity per se; instead, they would accept and even make an inequitable allocation if it can be justified.

Shaw and colleagues provided an impartiality account for this self-disadvantaging phenomenon (Shaw, 2013; Shaw et al., 2016). Acting impartially means not favoring any agent, especially not benefiting oneself if his (or her) self-interest is involved, which serves to avoid being judged negatively by recipients or potential observers (Choshen-Hillel et al., 2015; DeScioli & Kurzban, 2013; Shaw, 2013). First-party self-disadvantaging allocators sacrificed their own benefit but promoted others' benefit, thus observed the principle of impartiality and achieved efficiency; third-party equitable allocators did not favor either recipient. Thus, both kinds of behaviors signal impartiality (Choshen-Hillel et al., 2015; Shaw, 2013; Shaw et al., 2016). Moreover, this self-disadvantaging behavior could not be fully explained by generosity or benevolence because older children as first parties were less likely than younger ones to make such a disadvantageous inequity if the extra reward has to be allocated among two other peers instead of one peer (Shaw et al., 2016). As children grew older, they became more concerned for appearing impartial (Shaw et al., 2016) and managing a prosocial reputation (Rapp, Engelmann, Herrmann, & Tomasello, 2019). In so doing, children need to infer how others would think about their behaviors and execute what they should do finally. Thus, the development of the self-disadvantaging behavior with age is probably associated with the improved socio-cognitive and cognitive capacities.

*Potential cognitive mechanisms*

The evidence mentioned above indicated that as age increased, children would become adult-like in solving allocation problems. In regards to potential cognitive mechanisms, theory of mind (ToM), the ability to reason about others' mental states and predict others' behaviors accordingly (Premack & Woodruff, 1978), has been argued to be conducive to fair allocations (e.g., Sally & Hill, 2006). Preschoolers with improved ToM were more likely to consider inequality unacceptable than those without (Mulvey, Buchheister, & McGrath, 2016). Many developmental psychologists typically employed two economic exchange games to study the role of ToM in the development of children's fairness-related behaviors, the Ultimatum Game (UG) and the Dictator Game (DG). In the UG, an anonymous proposer needs to propose an allocation plan and an anonymous responder can accept or reject it (Camerer, 2003). If the responder accepts the plan, the resource will be distributed as proposed; otherwise, neither of them will get any resource. Thus, the proposer needs to infer the desire of the responder in order to make the proposal accepted. The DG is a bit different in that the responder has to accept the proposal (Camerer, 2003). Previous studies found that ToM could positively predict children's fair behaviors in the UG (Sally & Hill, 2006; Takagishi, Kameshima, Schug, Koizumi, & Yamagishi, 2010) as well as the DG (Sally & Hill, 2006; Wu & Su, 2014; Yu, Zhu, & Leslie, 2016). Given the definition of ToM, it could be assumed that ToM might help children better understand what kind of allocations the recipients would desire if the recipients have the right to reject it in the UG ("she may want me to split the rewards, or she will reject what I propose"), or how recipients and potential observers in the DG would think about their allocation behaviors ("she will be happy if I give half to her").

Similarly, in a context where resources cannot be allocated equitably, we supposed that preschoolers with higher levels of ToM might infer what others would consider socially desirable. If children's own interest was involved, they would reason that "she will be happy if I give more to her and this can save the resource", knowing that giving more to the other is preferable; if not, they would reason that "she will be sad if she receives less than the other one", knowing that wasting the extra reward is preferable. Hence, we supposed that ToM would be positively associated with preschoolers' self-disadvantaging behaviors.

Moreover, in dealing with fairness issues, one usually needs to find a balance between the desire to maximize his or her own interest and a motivation to conform to social norms (e.g., Sanfey, Rilling, Aronson, Nystrom, & Cohen, 2003). Inhibitory control (IC), an ability to suppress prepotent responses when pursuing a cognitively represented goal (Diamond, 2013), is positively related to fair behaviors (e.g., Blake, Piovesan, Montinari, Warneken, & Gino, 2015). Some studies also demonstrated a causal relation between IC and fairness-related behaviors. Steinbeis and Over (2017) found that children shared more rewards with anonymous peers in the DG after their IC was temporally enhanced by listening to a story in which the protagonist exerted IC to resist temptation than after listening to a neutral control story; and this priming effect could not be explained by perceptions of fairness. Similarly, Steinbeis (2018) had a group of children complete a Stop-signal-reaction-time task before allocation task to lower their IC temporally, and found that this experiment group shared less rewards with peers in the subsequent DG than the control group did. These findings suggest that IC plays a crucial role in inhibiting one's selfish impulse and aligning one's behavior with prosocial norms. Thus, we proposed that IC would be associated with a tendency to divide resources in an unselfish way when children have to make a trade-off between equity and efficiency.

To achieve the cognitive goal of meeting social norms or what is socially desirable, in addition to inhibiting one's selfish impulse, IC could also play a role in suppressing other internal predispositions like envy (Rawls, 1971; Takahashi et al., 2009) or concern for efficiency (Engelmann & Strobel, 2004; Rawls, 1971). Children would not only focus on the absolute gains, but also their relative gains compared with peers (Blake et al., 2014). Thus, in Shaw and colleagues' study (2016), giving such a disadvantageous inequity allocation demanded inhibiting one's envious impulse and social comparison preference (Fehr & Schmidt, 1999; Festinger, 1954; Sznycer et al., 2017; Takahashi et al., 2009), especially when children as first-parties could choose only between making a self-disadvantageous allocation or an equitable allocation. And when a third-party allocates resources between two equally deserving, anonymous recipients, making an equitable allocation means overriding one's concern for improving efficiency of allocation in the dilemma of equity-efficiency conflict. Thus, IC should play a substantial role in inhibiting these predispositions if a person wants to behave impartially.

Neuroscience studies provided more evidence for such a role. Steinbeis and colleagues (2012) investigated decisions of 6- to 13-year-old children in the UG and DG as well as neural mechanisms of the process. The difference in offer size between UG and DG reflects the ability to reconcile one's own needs and social norms (Camerer, 2003; Steinbeis et al., 2012). They found that age and IC were positively correlated with the difference between amount allocated to the other in the UG and DG, but not correlated with their social norm understanding like fairness judgment. Moreover, the activation in dorsolateral prefrontal cortex (dlPFC), an area associated with IC (Miller & Cohen, 2001), as well as its cortical thickness was positively associated with IC and the difference in offer size between UG and DG (Steinbeis et al., 2012). The evidence indicated that IC helps individuals narrow the gap between their knowledge of social norms and actual behaviors. In other words, IC would help align behaviors with children's knowledge about fairness. Thus, children would demonstrate more self-disadvantaging behaviors as age increased because their IC improved with age.

Thus, ToM and IC are potential cognitive basis of children's self-disadvantaging behaviors. Specifically, ToM enables children to infer the recipient's social preference, and IC would help children inhibit their selfishness and envy. We chose 4- to 6-year-old preschoolers for two reasons. First, preschoolers' understanding of equity and impartiality developed fast during the preschool period. For example, 5-year-olds make more equitable distributions than 4-year-olds (Lane & Coon, 1972), and 5-year-olds also could understand procedural justice and accept an inequitable distribution made through an impartial procedure (Grocke et al., 2015). Second, their ToM and IC also develop fast. Children from about 4 years old on could pass the false-belief task, which meant that they could reason about others' beliefs (Wellman, Cross, & Watson, 2001). IC also began to develop rapidly at age 4 with behavioral improvement and functional changes in the neural substrates (Davidson, Amso, Anderson, & Diamond, 2006).

*The current study*

In summary, the current study was designed to investigate the cognitive mechanisms underlying the development of the self-disadvantaging behavior in the preschool period. Findings were expected to give a cognitive account of such age-related difference in self-disadvantaging behaviors and shed light on how to promote children's actual fairness-related behaviors.

We employed a real-life-like resource allocation task designed specifically for young children. We randomly assigned 4- to 6-year-old children into the first-party or the third-party conditions in Experiment 1. In the first-party condition, children were asked to choose among creating a self-advantaging inequity (self-benefit), wasting one extra reward bell (equity), and creating a self-disadvantageous inequity (other-benefit); in the third-party condition, their decisions were not related to their own benefits, and they needed to allocate an odd number of toy bells between two equally deserving recipients. We hypothesized that in the first-party condition, the tendency to create a self-disadvantageous inequity would be positively associated with age, IC and ToM; in the third-party condition, the tendency to adhere to equity would be positively associated with age, IC and ToM. However, it was to be explored, to exhibit the self-disadvantaging behavior, whether one should only inhibit selfish impulses with IC or should also inhibit other tendencies like envy that prevented one from conforming to social norms. Thus, we conducted Experiment 2 to reveal the specific role of IC with fewer choices. Children in Experiment 2 could only choose between equity and other-benefit. Additionally, we proposed that the development of IC and ToM capabilities could explain the age-related differences in children's self-disadvantaging behaviors when this required inhibition of prepotent responses and reasoning how others would perceive the decisions.

# Experiment 1
## Method

*Participants*

A total of 101 children aged 4 to 6 participated in this study, but 2 subjects were excluded because they disliked the toy bells as rewards for allocation, resulting in a final sample of 99 ($M_{age}$ = 65.01 months, $SD$ = 9.35, range = 47.80 to 78.44; 55 girls). The sample was drawn from 2 private kindergartens in Beijing, China, which served mostly children from middle class families. All subjects were ethnic Chinese, were born and had lived in mainland China all or most of the time. They were randomly assigned to one of two conditions: the first-party condition ($N$ = 54, $M_{age}$ = 64.96 months, $SD$ = 9.04, range = 48.43 to 78.44; 29 girls) and the third-party condition ($N$ = 45, $M_{age}$ = 65.07 months, $SD$ = 9.81, range = 47.80 to 78.08; 26 girls). Each group had an even gender composition ($p$s > .100). Informed consent was obtained from the school authorities and parents.

*Procedure*

A female experimenter tested each participant individually in a sequence of four tasks, which took about 25min in total in a quiet room at the participant's kindergarten. Each subject was instructed to complete, first, a resource allocation task (4 min), then a training with the future probability ordering task for the preparation of measuring ToM (6 min), followed by an interpretive theory-of-mind probability task measuring ToM (10 min), and finally a day-night task measuring IC (5 min). Each child was awarded 2 bells for participation after finishing all the tasks.

*Resource Allocation Task.* We asked participants to help allocate the reward bells, which was adapted from previous studies (Shaw et al., 2016; Shaw & Olson, 2012).

To place children in a situation involving their self-interest, each child in the first-party condition was asked to complete a filler task (a game of feeding animals) under the instruction of the experimenter. Then the child was told by the experimenter that she would leave to check another child's performance. Later, the experimenter came back, and the participant was told to allocate 5 bells as reward. They could choose among allocating the fifth bell to self, no one and other. Children in the third-party condition were directly asked to choose to allocate the fifth bell to one of the two equally deserving recipients or to waste it (see details in the supplementary material). This one-shot play design would reduce the possibility of expecting future reciprocity (Fehr & Fischbacher, 2003).

*Future likelihood ordering task.* This task was employed to train each child to indicate estimated probabilities from 1 to 10 demonstrated by stacked bars in an arrow for subsequent assessment of ToM (Lagattuta, Sayfan, & Harvey, 2014; see details in the supplementary material).

*Interpretive Theory-of-Mind Probability (ITomP) Task.* The ITomP task (Lagattuta et al., 2014) could be used to measure the mindreading ability for preschoolers. We adopted the "knowledgeable" condition of the task to test 4- to 6-year-olds' ToM instead of false belief tasks, which might be too easy for 5- to 6-year-olds (Lagattuta et al., 2014; Wellman et al., 2001). The task included 2 trials in total (the cloud and the castle trials). Participants were first shown the actual picture in each trial. And then they were asked to infer how a "naïve" agent would think what the actual picture was when he (or she) could only see an ambiguous section of the actual picture from a window (an arc for the cloud trial and a right angle for the castle trial). There were two indicators of this task, the inferred probability of actual pictures (a negative indicator of ToM) and prototypical pictures (a positive indicator of ToM).

*Day-Night Task.* In order to measure IC, the participant was asked to complete a Stroop-like task, i.e. day-night task (Lagattuta, Sayfan, & Monsour, 2011). It featured ten pictures of the sun and ten pictures of the moon. This version of task was better for the current study because classic versions might be too simple for older preschoolers (Lagattuta et al., 2011). The participant was asked to say the opposite name of each picture as quickly as possible ("day" for a moon, "night" for a sun). After answering four practice trials correctly, they would enter a formal test of twenty trials.

*Coding and Analysis*

Our study aimed to explore the underlying cognitive mechanisms of self-disadvantaging behaviors in preschoolers. For a cross-condition analysis, condition was coded into a binary variable (0 = third-party condition, 1= first-party condition), and allocation was first coded into a binary variable (0 = not to give to other, 1 = to give to other). Allocation to either recipient in the third-party condition would be both inequitable and unfair, yet in the first-party condition, allocation to the other would seem impartial. Thus, if allocation choice would be predicted by certain variables of interest, condition was expected to play a moderating role in such a relation. Notably, for the first-party condition, allocation decision was re-coded into an ordinal variable (0 = self-benefit, 1= equity, and 2 = other-benefit). Since the self-benefit choice was self-centered, selfish, and would be socially undesirable, so we coded it as 0; throwing the extra bell away preserved equity but failed to take efficiency into account, so we coded it as 1; the other-benefit choice would be both efficient and impartial, and according to previous studies, older children were more likely than younger children to choose other-benefit over equity (Shaw et al., 2016), similar to what adults did (Choshen-Hill et al., 2015), so we coded it as 2. In the IToMP task, the inferred probabilities of actual and prototypical pictures were selected as the previous study did (Lagattuta et al., 2014). They were recorded as scores ranging from 1 to 10. The accuracy rate of the IC task was recorded in percentage values.

We mainly used R 3.2.0 (R Core Team, 2013) to analyze the data. The *glm* function was employed for logistic regression analysis of children's choices across conditions (1 or 0: to give to other or not) (Swiss Federal Institute of Technology in Zurich, 2014a), the *multinom* function from the *nnet* package was employed for multinomial logit regression of children's choices in the first-party condition (self-benefit, equity, other-benefit) (Venables & Ripley, 2002), and the *polr* function from the *MASS* package was employed for ordinal logistic regression of the ternary outcomes in the first-party condition (0 to 2: self-benefit, equity, other-benefit) (Swiss Federal Institute of Technology in Zurich, 2014b). We estimated effect sizes by calculating odds ratios (*OR*s) and their 95% confidence intervals (CIs). We standardized the continuous variables before the analysis in order to make units comparable (Cohen, Cohen, West, & Aiken, 2003). To conduct a mediation analysis in order to investigate the cognitive mechanisms underlying the development of impartial behaviors, we used the Preacher and Hayes (2008) mediation setup for the binary outcome variable, whereas the method reported in Liu and colleagues' study (2015) for the ordinal outcome variable.

**Results**

*Preliminary analysis*

Zero-order correlations among focal variables were conducted for the first-party condition, the third-party condition and across the conditions (see Table 1S in the supplementary material). Children in the two groups did not differ in age, ToM or IC ($p$s > .100). There were either no gender differences in age, ToM, IC or inclination to give the extra bell to other ($p$s > .100). Thus, gender was not further analyzed.

*The development of impartial behaviors in preschoolers*

First, we conducted a binomial logistic regression in which children's choices (1 or 0: to give to other or not) served as the dependent variable, and age, condition along with the interaction between them served as the independent variables. The effect of the interaction was significant, $b = 1.21$, $SE = 0.49$, $z = 2.49$, $p = .013$, $OR = 3.35$, 95% CI [1.35, 9.17]. And the effects of age and condition on children's choices were also significant, $b = -0.73$, $SE = 0.35$, $z = -2.13$, $p = .034$, $OR = 0.48$, 95% CI [0.23, 0.90]; $b = -1.48$, $SE = 0.46$, $z = -3.23$, $p = .001$, $OR = 0.23$, 95% CI [0.09, 0.55], respectively; McFadden's $R^2 = .13$.

Considering the choices across conditions were not parallel, thus we analyzed the relationship between age and children's choices for each condition separately. For the first-party condition, we conducted an ordinal logistic regression in which children's choices (0 to 2: self-benefit, equity or other-benefit) served as the dependent variable, and age as the independent variable. The effect of age on children's choices was significant, such that older children were more likely to choose equity and other-benefit, $b = 0.70$, $SE = 0.28$, $t = 2.50$, $p = .012$, $OR = 2.01$, 95% CI [1.18, 3.57]; McFadden's $R^2 = .06$. The multinomial logistic regression for the first-party condition also showed that age could significantly predict children's choices, $\chi^2(2) = 8.28$, $p = .016$, McFadden's $R^2 = .07$. For the third-party condition, we conducted a binomial logistic regression in which children's choices (1 or 0: to give to other or not) served as the dependent variable and age as the independent variable. As children grew older, they were less inclined to create inequity in order to achieve impartiality, $b = -0.73$, $SE = 0.35$, $z = -2.13$, $p = .034$, $OR = 0.48$, 95% CI [0.23, 0.90]; McFadden's $R^2 = .09$.

In addition, we also examined whether children were choosing at random for each condition. For the first-party condition, Chi-square goodness-of-fit test showed that the proportion of 4-year-old children (11 out of 17) choosing self-benefit was significantly higher than the chance level (1/3), $\chi^2(2) = 8.94$, $p = .011$, while statistically, 5-year-olds (6 out of 19) and 6-year-olds (3 out of 18) chose self-benefit at chance level, $\chi^2(2) = 0.74$, $p = .692$ and $\chi^2(2) = 3.00$, $p = .223$, respectively (see the left panel of Figure 1). For the third-party condition, the Chi-square goodness-of-fit test showed that only the number of 6-year-old children (9 out of 14) choosing equity (waste the extra bell) was significantly at above chance level (1/3), $\chi^2(1) = 6.04$, $p = .014$, while statistically, 4-year-olds (2 out of 12) and 5-year-olds (7 out of 19) chose equity at chance level, $\chi^2(1) = 1.50$, $p = .221$ and $\chi^2(1) = 0.11$, $p = .746$, respectively (see the right panel of Figure 1).

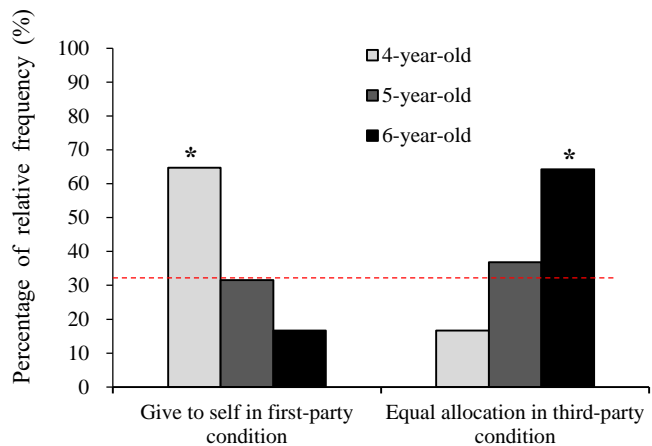

**Figure 1.** Proportion of children choosing self-benefit in the first-party condition or equity in the third-party condition, presented separately for age. $^*$ $p < 0.05$

*The underlying mechanisms of impartial behaviors in preschoolers*

As the correlation analysis implied, preschoolers' impartial behaviors were positively correlated with their IC and ToM. Specially, their choices were significantly correlated with the accuracy rate of IC task and the inferred probability of prototypical pictures in IToMP task. Hence, we performed regression analysis using these indicators. First, we conducted a binomial logistic regression, in which children's choices (1 or 0: to give to other or not) served as the dependent variable, and age, condition, the accuracy rate of IC task, the inferred probability of prototypical pictures, the interaction between the accuracy rate of IC task and condition (1 = the first-party condition, 0 = the third-party condition), plus the interaction between the inferred probability of prototypical pictures and condition served as the independent variables. The results showed that the interaction between condition (First-party) and the accuracy rate of IC task was significant, $b = 4.22$, $SE = 1.06$, $z = 3.98$, $p < .001$, $OR = 68.01$, 95% CI [10.30, 678.20]; the effects of condition and the accuracy rate of IC task on children's choices were also significant, $b = -2.73$, $SE = 0.74$, $z = -3.71$, $p < .001$, $OR = 0.07$, 95% CI [0.01, 0.24]; $b = -1.75$, $SE = 0.72$, $z = -2.42$, $p = .016$, $OR = 0.17$, 95% CI [0.03, 0.60], respectively. However, condition did not show a significant moderating effect on the relation between the inferred probability of prototypical pictures and children's choices, $b = 0.54$, $SE = 0.51$, $z = 1.06$, $p = .288$, $OR = 1.71$, 95% CI [0.63, 4.70] (see Table 1); McFadden's $R^2 = .34$.

**Table 1**
Results of logistic regression on giving to other or not for Experiment 1

| Variables | $b$ | $SE$ | $z$ | $p$ | $OR$ | 95% CI |
|---|---|---|---|---|---|---|
| (Intercept) | 1.11 | 0.51 | 2.19 | .028 | 3.03 | [1.27, 9.78] |
| Age (months) | -.29 | 0.36 | -0.80 | .422 | 0.75 | [0.37, 1.51] |
| Condition (First-party) | -2.73 | 0.74 | -3.71 | < .001 | 0.07 | [0.01, 0.24] |
| IC accuracy | -1.75 | 0.72 | -2.42 | .016 | 0.17 | [0.03, 0.60] |
| Prototypical pictures | -.32 | 0.36 | -0.89 | .375 | 0.73 | [0.35, 1.47] |
| Condition (First-party): IC accuracy | 4.22 | 1.06 | 3.98 | < .001 | 68.01 | [10.30, 678.20] |
| Condition (First-party): Prototypical pictures | 0.54 | 0.51 | 1.06 | .288 | 1.71 | [0.63, 4.70] |

We analyze the contribution of IC and ToM to preschooler's impartial behaviors for each condition separately. For the first-party condition, we conducted a multinomial logistic regression and an ordinal logistic regression in which age, the accuracy of IC task, the inferred probability of prototypical pictures served as the independent variables, and children's choices served as the dependent variable. The results of multinomial logistic regression showed that only the effect of IC on children's choices was significant, $\chi^2(2) = 17.10$, $p < .001$, McFadden's $R^2 = .24$ (see Table 2 for comparisons based on log odds). The results of ordinal logistic regression showed consistent evidence that the accuracy rate of IC task could only significantly predict children's choices after controlling for age, $b = 1.20$, $SE = 0.40$, $t = 3.03$, $p = .002$, $OR = 3.31$, 95% CI [1.63, 7.82], while the effects of age, the inferred probability for prototypical pictures were not significant ($ps > .05$); McFadden's $R^2 = .17$. For the third-party condition, the results of binomial logistic regression showed that the accuracy rate of IC task could only significantly predict children's choice of not wasting after controlling for age, $b = -1.76$, $SE = 0.74$, $z = -2.39$, $p = .017$, $OR = 0.17$, 95% CI [0.03, 0.61], but the effects of other predictors were not significant ($ps > .05$); McFadden's $R^2 = .27$.

**Table 2**

Multinomial logistic regression for children's choices in the first-party condition

| Independent variables | Comparisons based on log odds | | |
|---|---|---|---|
| | Equity vs. Self-benefit | Other-benefit vs. Self-benefit | Other-benefit vs. Equity |
| Age (months) | 0.36 | -0.16 | -0.52 |
| IC accuracy | 0.37 | 2.69* | 2.32* |
| Prototypical pictures | 0.88 | 0.92 | 0.04 |

Note: * $p < .01$. Self-benefit means children giving the extra bell to self, equity means children wasting the extra bell, and other-benefit means children giving the extra bell to other.

In order to explain the developmental trajectory of impartial behaviors during the preschool period, we tested the mediation effect of IC on the relation between age and children's choices for each condition. For the first-party condition, age served as an independent variable ($X$), the accuracy rate of IC task served as a mediator ($M$), and children's choices (0 to 2: self-benefit, equity or other-benefit) served as a dependent variable (*Y*). We found that as age increased, children were more likely to choose equity and other-benefit (*c* = 0.70, *SE* = 0.28, *t* = 2.50, *p* = .012, *OR* = 2.01, 95% CI [1.18, 3.57]), and their accuracy rate of IC task increased (*a* = 0.64, *SE* = 0.13, *t* = 4.99, *p* < .001). Moreover, children with higher levels of IC had a stronger tendency for equity and other-benefit choices (*b* = 1.18, *SE* = 0.40, *t* = 2.96, *p* = .003, *OR* = 3.24, 95% CI [1.59, 7.68]), but the direct effect was not significant now, *c'* = 0.15, *SE* = 0.33, *t* = 0.46, *p* = .649, *OR* = 1.16, 95% CI [0.60, 2.24] (see the top panel of Figure 2). The indirect effect of age on children's choices in the first-party condition through IC was significant, *ab* = 0.75, 95% CI [0.17, 1.32] (Liu, Zhang, & Luo, 2015).

Likewise, for the third-party condition, age served as an independent variable (*X*), the accuracy rate of IC task served as a mediator (*M*), and children's choices (1 or 0: to give to other or not) served as a dependent variable (*Y*). The results showed that as age increased, children were less likely to give the extra bell to one of the recipients (*c* = -0.73, *SE* = 0.35, *z* = -2.13, *p* = .034, *OR* = 0.48, 95% CI [0.23, 0.90]), and the accuracy rate of IC task increased (*a* = 0.40, *SE* = 0.11, *t* = 3.55, *p* = .001). Moreover, children with higher levels of IC were less inclined to give the extra bell to the recipients (a stronger tendency to equitable allocations) (*b* = -1.85, *SE* = 0.72, *z* = -2.59, *p* = .010, *OR* = 0.16, 95% CI [0.03, 0.53]), but the direct effect was not significant now, *c'* = -0.33, *SE* = 0.43, *z* = -0.75, *p* = .453, *OR* = 0.72, 95% CI [0.29, 1.90] (see the bottom panel of Figure 2). The indirect effect of age on children's choice in the third-party condition through IC was significant, *ab* = -0.75, 95% CI [-2.22, -0.11] (Preacher & Hayes, 2008). Thus, the results across conditions showed that the accuracy rate of IC task played a fully mediator in the relation of age on preschoolers' impartial behaviors.

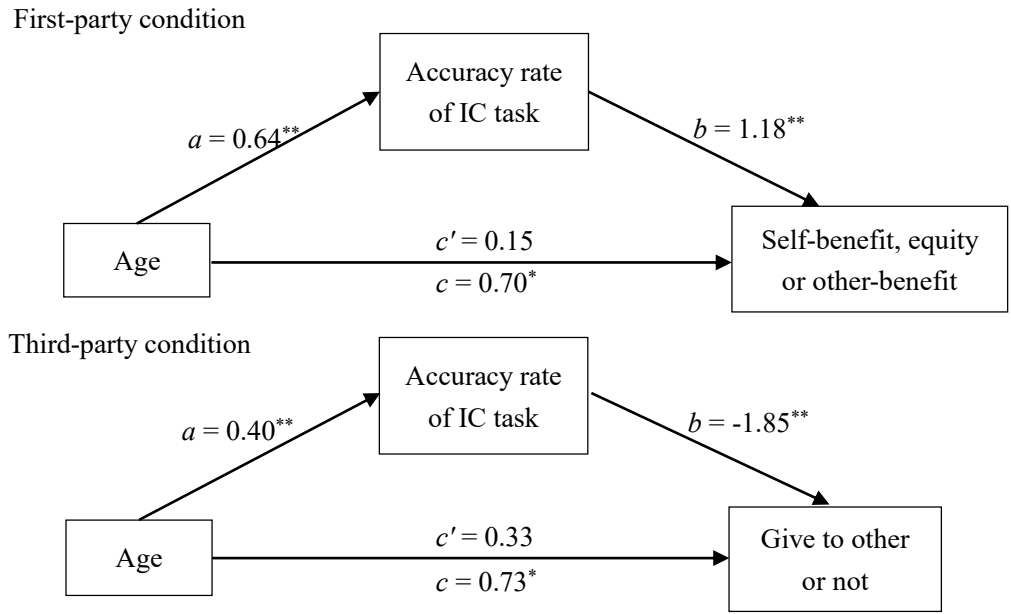

**Figure 2.** IC mediated the effect of age on children's choices in different conditions. The mediation effects of IC in both conditions were tested separately. Results showed that IC played a full mediating role in the relation between age and children's choices in both conditions. Note: $^*p < .05$, $^{**}p \leq .01$.

## Discussion

The current results of the self-disadvantaging phenomenon and its development replicated findings from previous studies (Choshen-Hillel et al., 2015; Shaw et al., 2016). There were asymmetric developmental trajectories in children's impartial behaviors across contexts. Six-year-old children did not show an unselfish tendency in the first-party condition whereas they showed a tendency for equity in the third-party condition.

More importantly, we found that IC was a full mediator for the age-related differences of impartial behaviors, while ToM might also make some contributions to it, but its particular role was not confirmed. It suggested that IC played a substantial role in implementing an impartial allocation, which was consistent with findings on children's fair behaviors obtained with other paradigms in previous studies (e.g., Aguilar-Pardo, Martínez-Arias, & Colmenares, 2013; Blake et al., 2015; Liu et al., 2016). The choices in Experiment 1, however, could not fully tell whether the role of IC was only for suppressing one's selfish impulses because children in the first-party condition could choose among self-benefit, equity or other-benefit. Whether IC would still play a role in children's self-disadvantaging behaviors when children were forced to choose between equity and other-benefit remained a question. As the definition of IC suggested, one has to suppress his (or her) impulses or habits in order to behave in a socially desirable way (Diamond, 2013). Individuals also had other impulses like envy. Thus we conducted Experiment 2 to find more evidence about how IC played a role in implementing this self-disadvantaging behavior.

## Experiment 2

In the context of equity-efficiency conflict, giving more to other when personal benefit was involved is a more socially desirable choice than wasting it (Choshen-Hillel et al., 2015; Shaw et al., 2016). Considering the three choices among which children could choose, we could not fully figure out the specific role of IC in implementing such a self-disadvantaging behavior. Whether IC played a role only in inhibiting one's selfishness or IC functioned as an executor to align one's behavior with social norms irrespective of different internal predispositions remained to be further examined. Thus, we employed the resource allocation task in the first-party condition of Experiment 1 with two limited choices, i.e., equity or other-benefit. The choices in Experiment 2 were more parallel to the third-party condition in Experiment 1 because in both experiments children needed to decide between allocating the extra bell to other or wasting it. The aim of Experiment 2 was to investigate to what extent preschoolers' self-disadvantaging behaviors were associated with IC for inhibiting envious impulses and ToM for inferring which choice was more socially desirable. We hypothesized that higher IC and ToM would still be positively associated with more self-disadvantaging decisions.

**Method**

*Participants*

G*Power (version 3.1) (Faul, Erdfelder, Buchner, & Lang, 2009) showed that a sample size of 41 was required for a power (1-β) of .80 to detect an effect of $f^2 = .20$ at α = .05. We thus recruited another 41 children ($M_{age}$ = 59.63 months, $SD$ = 5.26, range = 48.33 to 70.63; 23 girls) coming from the same kindergartens as Experiment 1 participated Experiment 2. All subjects liked the reward bells. They were ethnic Chinese, were born and had lived in mainland China all or most of the time. Informed consent was obtained from the school authorities and parents.

*Procedure*

The procedures in the self-disadvantaging condition of Experiment 2 were similar to those in the first-party condition of Experiment 1, except that the "giving to self" option was removed in the resource allocation task. The order of the choices of giving more to other and wasting the extra reward in narration was counterbalanced between subjects. After completing the allocation task, participants were instructed to complete the measurement of ToM and IC as Experiment 1.

**Results**

Data of 4- and 5-year-olds from the third-party condition of Experiment 1 and from the self-disadvantaging condition of Experiment 2 were compared to control the effect of age. For this pooled sample of 72 ($N_{\text{Exp.1 third-party}} = 31$, $N_{\text{Exp.2}} = 41$), zero-order correlations among variables in the self-disadvantaging condition of Experiment 2 were conducted (see Table 2S in the supplementary material).

We conducted a binomial logistic regression, in which children's choices (1 or 0: self-disadvantaging or equity) served as the dependent variable, and age, condition, the accuracy rate of IC task, the inferred probability of prototypical pictures, the interaction between the accuracy rate of IC task and condition (1 = the self-disadvantaging condition in Experiment 2, 0 = the third-party condition in Experiment 1), plus the interaction between the inferred probability of prototypical pictures and condition served as the independent variables. Results showed that the interaction between condition (Self-disadvantaging condition) and the accuracy rate of IC task was significant, $b = 2.62$, $SE = 0.96$, $z = 2.73$, $p = .006$, $OR = 13.76$, 95% CI [2.62, 119.64]; the effect of the accuracy rate of IC task on children's choices was also significant, $b = -1.78$, $SE = 0.88$, $z = -2.02$, $p = .043$, $OR = 0.17$, 95% CI [0.02, 0.73]. However, condition did not show a significant moderating effect on the relation between the inferred probability of prototypical pictures and children's choices (1 or

0: to give to other or not), $b = 0.46$, $SE = 0.60$, $z = 0.76$, $p = .448$, $OR = 1.58$, 95% CI [0.49, 5.48] (see Table 3); McFadden's $R^2 = .21$.

**Table 3**
Results of logistic regression on self-disadvantaging or equity for Experiment 2

| Variables | b | SE | z | p | OR | 95% CI |
|---|---|---|---|---|---|---|
| (Intercept) | 1.25 | 0.60 | 2.06 | .039 | 3.47 | [1.21, 14.15] |
| Age (months) | -0.06 | 0.34 | -0.18 | .859 | 0.94 | [0.48, 1.83] |
| Condition (Self-disadvantaging) | -1.32 | 0.72 | -1.85 | .065 | 0.27 | [0.05, 0.98] |
| IC accuracy | -1.78 | 0.88 | -2.02 | .043 | 0.17 | [0.02, 0.73] |
| Prototypical pictures | -0.55 | 0.48 | -1.16 | .248 | 0.58 | [0.20, 1.42] |
| Condition (Self-disadvantaging): IC accuracy | 2.62 | 0.96 | 2.73 | .006 | 13.76 | [2.62, 119.64] |
| Condition (Self-disadvantaging): Prototypical pictures | 0.46 | 0.60 | 0.76 | .448 | 1.58 | [0.49, 5.48] |

We analyze the contribution of IC and ToM to preschooler's impartial behaviors for each condition separately. The results of the binomial logistic regression analysis for the self-disadvantaging condition showed that the accuracy rate of IC task could only significantly predict children's self-disadvantaging choice after controlling for age, $b = 0.98$, $SE = 0.46$, $z = 2.12$, $p = .034$, $OR = 2.68$, 95% CI [1.19, 7.66], while the effects of age, the inferred probability for prototypical pictures were not significant ($ps > .05$); McFadden's $R^2 = .14$ (see the left panel of Figure 3). For the third-party condition, similarly, the accuracy rate of IC task could only significantly predict children's equity choice after controlling for age, $b = -1.91$, $SE = 0.92$, $z = -2.09$, $p = .037$, $OR = 0.15$, 95% CI [0.02, 0.66], but the effects of other predictors were not significant ($ps > .05$); McFadden's $R^2 = .28$ (see the right panel of Figure 3).

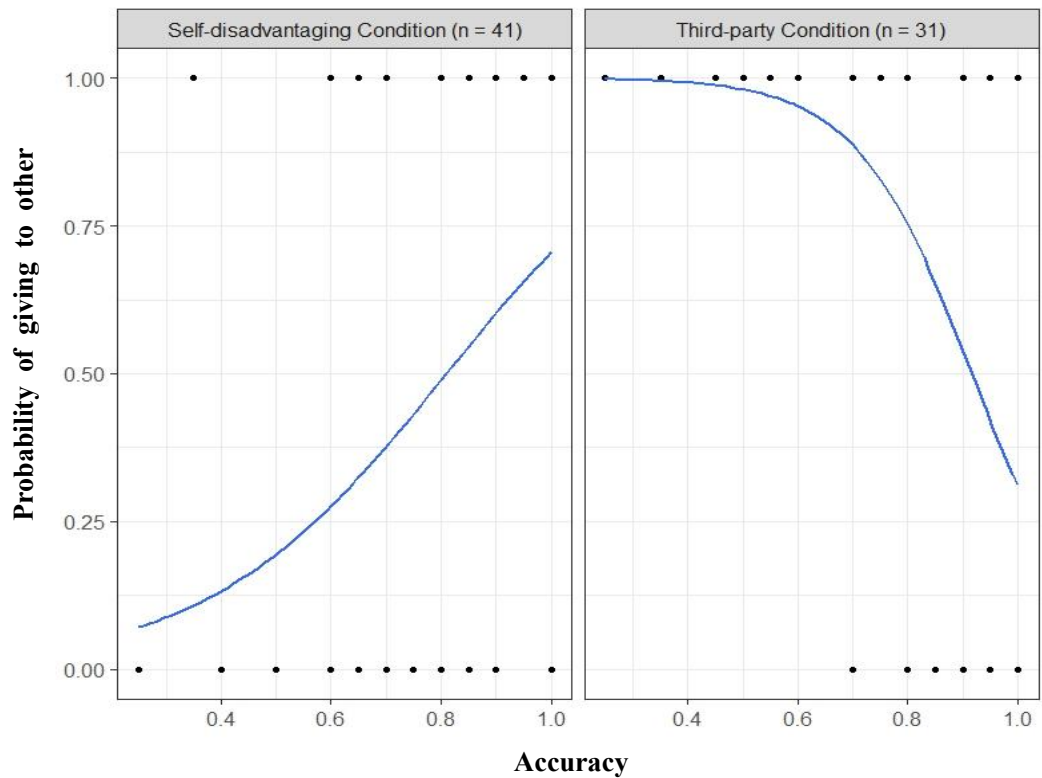

**Figure 3.** Binomial logistic regression of accuracy of the IC task on probability of giving to other in each condition.

**Discussion**

In Experiment 2, children were faced with limited choices between creating self-disadvantageous inequity and wasting the extra piece of resource. We found that IC was still positively predictive of their self-disadvantaging choices. This result eliminated the possibility that IC only played a role in overriding one's selfishness. Instead, it would also serve to inhibit one's envy in allocation. We speculated that preschoolers might perceive giving the extra bell to other as more socially desirable than wasting it if their self-interest was involved (Choshen-Hillel et al., 2015; Shaw et al., 2016) because some children could give justified reasons like "I don't want to get the bell rusted" or "then I will yield it to her" as they talked to themselves when making the decisions.

This self-disadvantaging phenomenon had something in common with previous findings with other paradigms (e.g., Grocke et al., 2015; Schmidt, Svetlova, Johe, & Tomasello, 2016), i.e., children's understanding of fairness is relatively sophisticated because they would consider inequity with justified reasons as fair. They would not only accept inequitable distributions made through an impartial lottery wheel instead of a partial one (Grocke et al., 2015), but also make inequitable distributions with legitimate reasons instead of idiosyncratic reasons such as "I just want more" (Schmidt et al., 2016). They would even make a self-disadvantageous inequity for efficiency as Shaw and colleagues' study (2016) as well as the current study suggested. All these findings indicated that children would display social behaviors consistent with social expectations (House, 2018), and IC would help to fill the gap between what is socially desirable and their actual performance (Diamond, 2013; Liu et al., 2016; Steinbeis et al., 2012)

**General Discussion**

The present study examined the cognitive mechanisms underlying the development of self-disadvantaging behaviors among preschoolers. In line with previous findings (Shaw et al., 2016), we replicated that older children in both the first-party and third-party conditions exhibited a stronger tendency than younger ones to make impartial decisions (some cultural and social-economic differences of the self-disadvantaging phenomenon were discussed in the supplementary material). Moreover, impartial behaviors increased with the development of IC and ToM. We also found that IC could explain the age-related differences in children's impartial behaviors where IC fully mediated the relation between age and impartial behaviors in both the first-party and third-party conditions.

*The role of IC*

We found that children with higher levels of IC tended to make inequitable yet impartial allocations, which disadvantaged themselves, instead of wasting the extra piece of resource. In Experiment 1, children in the first-party condition had three choices in total: self-benefit, equity or other-benefit. As Sanfey and colleagues (2003) argued, one had to inhibit his (or her) selfish impulse before he or she would decide not to choose "the self" option, because one had to make a balance between maximizing economic self-interest, and conforming to the social norm, and "inhibition" of selfishness was required for the latter.

Selfishness, however, is not the only form of prepotent response to inhibit if one is to meet the demand of social norms in social interactions. In Experiment 2, children had to choose between giving more to other (other-benefit) or wasting the extra bell (equity). Previous studies showed that children had a strong inequity aversion, especially for disadvantageous inequity (Blake & McAuliffe, 2011). In Fehr and colleagues' study (2008), most 5- to 6-year-olds chose the equal allocation (1:1) instead of the self-disadvantaging allocation (1:2). Such a strong tendency for egalitarianism might be associated with envy (Dawes, Johnson, Smirnov, Fowler, & McElreath, 2007; Sznycer et al., 2017), which was a dark side of human nature (Smith & Kim, 2007). According to the definition of envy (Smith & Kim, 2007), getting fewer desired rewards (note that all the children recruited in the current study liked the bells very much) than another person for equal performance would evoke feelings of inferiority or resentment. Thus, unlike just providing two choices (1:1 or 1:2) in Fehr and colleagues' study (2008), when emphasizing the possibility of wasting the extra bell, children in both Experiments 1 and 2 needed to suppress their impulses of envying the other getting more desired reward for equal effort if they chose to give more to other.

The current study extended previous findings that IC played a critical role in compliance with equality in the DG (Blake et al., 2015) and enforcing fair norms (Steinbeis, 2018) as children grew older. We speculated that IC also contributed to impartiality including achieving equity at the expense of wasting some extra resource as third-parties and making a self-disadvantageous inequity as first-parties. Taken together, it suggested that IC played a substantial role in implementing fair decisions, including inhibiting one's selfishness and other prepotent responses like envy.

Notably, the current format of the resource allocation task is essential for capturing the effect of IC on fair behaviors. This format ensured whether children decided to disadvantage themselves, keep equity, or disadvantage others, they had to make one active (motor) choice. In contrast, in Smith and colleagues' study (2013), children in the resource allocation task did not need to do anything if they wanted to disadvantage others. In resource allocation tasks, asking children which allocation (e.g., 4:0, 3:1, 2:2) they want to choose may be more likely to detect the effect of IC than asking children how many items they want to give away, because the former format involves inhibiting a motor response such as choosing the 4:0 allocation if children want to behave prosocially. Thus, future studies in this area should make all the choices parallel.

*The role of ToM*

The present study found that ToM was positively correlated with children's impartial behaviors in the resource allocation task in Experiment 1 but failed in Experiment 2. There might be two possible explanations. Firstly, ToM contributes to the process of overriding self-interest (Yu et al., 2016), but not likely to the process of suppressing the envious responses. Children in previous studies employing UG or DG just needed to consider whether to favor themselves or to conform to social norms, which was related with their ToM (e.g., Wu & Su, 2014). However, creating a disadvantageous inequity by themselves might rely more on inhibition rather than mentalizing, as shown by the insignificant relation between ToM and self-disadvantaging choices in Experiment 2 and the insignificant mediation effect of ToM on the relation between age and impartial behaviors in Experiment 1. Participants would struggle a lot between giving more to self, no one and other. They might try to infer how others would think about each choice they were to make in the resource allocation task, but the final formation of choice depended on IC instead of ToM, which was also supported by Steinbeis and colleagues' study (2012). Thus, future studies need to further examine the neural mechanisms or the time course of impartial decision making with other technologies, such as neural oscillation through EEG.

Another possibility was that ToM did play a role in impartial behaviors though there were only two choices, i.e., equity or other-benefit. But the IToMP task in the current study might have failed to capture certain aspects of ToM, as ToM was a multi-dimensional construct, which consisted of intention, emotion, desire, knowledge, etc. (Wellman, 2002). Individuals' fair behaviors were the result of a dual process involving emotion and cognition (Beugré, 2009). The IToMP task used in the present study only reflected children's abilities to infer others' knowledge state, which was a cognitive component of ToM.

*Contributions and limitations*

The current study examined the cognitive mechanisms underlying the development of self-disadvantaging behaviors over the preschool period. The findings of the current study showed that IC played a critical role in the implementation of self-disadvantaging behaviors when this implementation required control over selfishness and envy, and ToM might also have some influence by promoting children's inference of others' social preference. The self-disadvantaging behavior is at least partially attributed to impartiality, a form of fairness (Choshen-Hillel et al., 2015; Shaw, 2013). Previous studies also suggested a role of IC as well as ToM in other forms of fairness, like equity (e.g., Wang & Su, 2013) or equality (e.g., Blake et al., 2015; Takagishi et al., 2010). All the findings suggested that humans' understanding of different forms of fairness might share the same cognitive mechanisms. The development of individuals' behaviors that met social expectations is largely correlated with their improvement of IC, and probably associated with the maturation of prefrontal cortex (Steinbeis et al., 2012). As age increases, children become better in IC, which helps them endorse fairness more to behave adaptively in social interactions.

However, the current study suffered from some limitations. First, it employed a correlational design to examine the relationship between IC and children's self-disadvantaging behaviors, and future research could employ manipulations like taxing cognitive control (Steinbeis, 2018) or priming (Steinbeis & Over, 2017) and test a potential causal relationship. Second, future studies could also employ other paradigms such as story tasks (e.g., Mills & Keil, 2008) to study children's developing notions of impartiality and compare findings obtained with different paradigms on the developmental trajectory of impartial behaviors.

*Practical implications*

Although some previous studies emphasized the important role of social-cognitive abilities in promoting children's prosociality (e.g., Sally & Hill, 2006), our findings as well as other studies (e.g., Steinbeis et al., 2012) highlight the importance of improving children's executive function in morality educational programs. To behave prosocially, inferring others' desires or reasoning how others think about the decision was important. Other processes such as inhibiting undesirable responses, regulating negative emotions, and making flexible decisions to meet the demand of specific situation, which relied largely on executive control, were also essential to the actualization of prosocial behaviors (Schonert-Reichl et al., 2015). Thus, instead of only telling children what is good and what is right, teachers could also include training programs such as those for children's inhibitory control or socio-cognitive abilities to promote children's morality in social interactions.

## Conclusion

To make a relatively satisfactory decision in a real-life dilemma where equity is in conflict with efficiency, preschoolers would be more likely to make a self-disadvantageous inequitable decision with self-interest involved as age increases. IC, which matures with age, could be a more direct determinant of the age-related differences. It plays a substantial role in implementing this self-disadvantaging behavior because it helps to inhibit children's impulses like selfishness and envy in order to behave in a more socially desirable way.

# Supplementary material

**Method**

*Resource Allocation Task*

Each child in the first-party condition was asked to complete a filler task (a game of feeding animals) under the instruction of the experimenter. Then the child was told by the experimenter that she would leave to check another child's performance. Later, the experimenter came back, and the participant was told:

Congratulations! You and a girl (or boy) named Hua (or Ming) each did a good job and got the same score in the game. And now we want to give you two some bells as a prize. Look, do you like these bells? (Show a total of 5 bells to the child.) But I do not know how many to give each of you. Could you help me? (OK.)

You will decide how many bells each of you will get. There are five. We have one for you, one for Hua (or Ming), one for you, and one for Hua (or Ming). Oh, there is one left. Do you think it should be given to Hua (or Ming), yourself, or no one? [The participant was asked to place the last bell into one of the three boxes with labels on their own. The order of "Hua (or Ming)" and "yourself" in the narration was counterbalanced between subjects.] If you decide to give it to no one, the bell will be put away and rust, which means no one can play with it.

The participant in the third-party condition were directly asked to help how to allocate the fifth bell. The experimenter told:

Just now, two children named Hua (or Ming) and Hong (or Mao) completed a game to feed animals. They each did a good job and got the same score. Thus we want to give them some bells

as a prize. But I don't know how many to give each of them. Could you help me? (OK.)

You will decide how many bells each of them will get. There are five. We have one for Hua (or Ming), one for Hong (or Mao), one for Hua (or Ming), and one for Hong (or Mao). Oh, there is one left. Do you think it should be given to Hua (or Ming), Hong (or Mao), or no one? If you decide to give it to no one, the bell will be put away and rust, which means no one can play with it.

"Hua" and "Hong" are common Chinese names for girls, while "Ming" and "Mao" are common Chinese names for boys. And the order of "Hua (or Ming)" and "Hong (or Mao)" in the narration was counterbalanced between subjects. Matching genders would prevent a gender preference from becoming a confounding variable (Renno & Shutts, 2015).

*Future likelihood ordering task*

This task was employed to train each child to indicate estimated probabilities from 1 to 10 demonstrated by stacked bars in an arrow for subsequent assessment of ToM (Lagattuta, Sayfan, & Harvey, 2014). The scale was subdivided into four sections: *definitely will not* (Bar 1), *might* (Bars 2 to 5), *probably will* (Bars 6 to 9), and *definitely will* (Bar 10). The participant learned an example trial for training. In the training trial, the experimenter showed four 3 × 3 in. pictures randomly placed in a 2 × 2 grid next to the arrow scale on the table, each showing a bucket filled with water to about half the depth and tilting at varying angles. The participant, with the help of the experimenter, placed each picture adjacent to the matching section under the arrow scale in order of increasing probability of the water spilling out from the bucket to the right, as the arrow pointed to. Then the participant took at least two similar test trials independently. In the event that the participant failed one of the two test trials, a third test trial would be added. All the participants either passed the first two test trials or made the third right and proceeded to the following task.

## Results

**Table 1S**

Correlation matrix of variables in Experiment 1

| Variables | Across conditions (N = 99) | | | | First-party condition (N = 54) | | | | Third-party condition (N = 45) | | | |
|---|---|---|---|---|---|---|---|---|---|---|---|---|
| | 2 | 3 | 4 | 5 | 2 | 3 | 4 | 5 | 2 | 3 | 4 | 5 |
| 1. Age (months) | -.06 | .40** | .52** | -.06 | -.10 | .44** | .57** | .31* | -.01 | .36* | .48** | -.33* |
| 2. Actual pictures | | -.27** | -.11 | .04 | | -.14 | -.05 | -.16 | | -.49** | -.24 | -.13 |
| 3. Prototypical pictures | | | .30** | -.04 | | | .18 | .27* | | | .43** | -.35* |
| 4. IC accuracy rate | | | | -.08 | | | | .57** | | | | -.51** |
| 5. Children's choices | | | | 1 | | | | 1 | | | | 1 |

Note: *$p < .05$ (two-tailed), **$p < .01$ (two-tailed). Children's choices were coded into a binary variable (1 or 0: to give to other or not) across the conditions and for the third-party condition, while an ordinal variable (0 to 2: self-benefit, equity or other-benefit) for the first-party condition.

**Table 2S**

Correlation matrix of variables in Experiment 2

| Variables | Across conditions (N = 72) | | | | Self-disadvantaging condition (N = 41) | | | | Third-party condition (N = 31) | | | |
|---|---|---|---|---|---|---|---|---|---|---|---|---|
| | 2 | 3 | 4 | 5 | 2 | 3 | 4 | 5 | 2 | 3 | 4 | 5 |
| 1. Age (months) | .09 | .25* | .40** | -.05 | .26 | .20 | .35* | .02 | -.08 | .35† | .46** | -.15 |
| 2. Actual pictures | | -.34** | .02 | .07 | | -.11 | .26 | -.06 | | -.59** | -.29 | .20 |
| 3. Prototypical pictures | | | .10 | -.26* | | | -.15 | -.10 | | | .36* | -.34† |
| 4. IC accuracy rate | | | | .03 | | | | .37* | | | | -.45* |
| 5. Self-disadvantaging or equity | | | | 1 | | | | 1 | | | | 1 |

Note: †$p < .10$ (two-tailed), *$p \leq .05$ (two-tailed), **$p < .01$ (two-tailed). Children's choices were coded into a binary variable (1 or 0: self-disadvantaging or equity) across the conditions, for the self-disadvantaging and third-party conditions.

**General discussion of cultural and social-economic differences**

Yet many studies about fairness-related behaviors and its development recruited participants from Westernized and industrialized nations like the United States (e.g., Shaw, Choshen-Hillel, & Caruso, 2016) and Germany (e.g., Schmidt, Svetlova, Johe, & Tomasello, 2016). Regarding to economic and cultural variations in human fairness (Blake et al., 2015; Henrich et al., 2010), how participants from an Eastern and developing country in a collective cultural background would behave is unclear.

The current findings of the self-disadvantaging phenomenon were consistent with previous studies (Choshen-Hillel et al., 2015; Shaw et al., 2016). It indicated that this phenomenon is fairly robust across cultures. But there might also be some cultural specificity in human fairness (Blake et al., 2015) and in resolution to conflicts between equity and efficiency. We compared the data in previous studies from different countries as below (see Table 3S). As the data suggested, children in a developed country with individualist culture (US) tended to waste the extra resource and uphold the equity principle, while children in developing countries with collectivist culture (China and Uganda) (Rarick et al., 2013) were more likely to give it to a non-self recipient. This descriptive analysis showed consistent results with those in previous studies, which indicated that the magnitude of inequity aversion varies in economic statuses and cultural norms (e.g., Henrich et al., 2010; Huppert et al., 2018). Individualist and developed societies emphasize independent effort or personal achievement (Berry, 1971), and might endow less weight on efficiency principle, especially when it is in conflict with equity principle. However, collectivist and developing societies concern much more about responsibility and group welfare (Ramamoorthy & Carroll, 1998). Thus, greater proportions of children in China and Uganda than US gave the extra reward to the other no matter with self-interest involved or not (Paulus, 2015; Shaw et al., 2016; Shaw & Olson, 2012). This implied that fairness is not solely an innate product of mind, but shaped by cultures and social norms (Henrich et al., 2010).

**Table 3S**

Comparison of impartial behaviors among preschoolers between different countries across conditions

| Country | First-party condition | | | | | Third-party condition | |
| --- | --- | --- | --- | --- | --- | --- | --- |
| | Two choices | | Three choices | | | | |
| | Waste | Other | Self | Waste | Other | Waste | Inequity |
| America (Shaw et al., 2016; Shaw & Olson, 2012) | 72% (47/65) | 28% (18/65) | 56% (46/82) | 20% (16/82) | 24% (20/82) | 60% (14/24) | 40% (10/24) |
| China (the current study) | 51% (21/41) | 49% (20/41) | 37% (20/54) | 35% (19/54) | 28% (15/54) | 40% (18/45) | 60% (27/45) |
| Uganda (Paulus, 2015) | — | — | — | — | — | 30% (10/33) | 70% (23/33) |